\documentclass[aps,amsfonts,floatfix,reprint,tightenlines,amssymb,superscriptaddress,twocolumn,longbibliography]{revtex4-2}  

\usepackage{amsmath}
\usepackage{subcaption}

\usepackage[linktocpage=true,
  colorlinks=true, 
  pdfborder={0 0 0},
  linkcolor=blue,
  citecolor=red,
  filecolor=yellow,
  urlcolor=blue,
  bookmarks,
]{hyperref}
\usepackage[capitalize]{cleveref}

\usepackage{orcidlink}

\newcommand{\angstrom}{\textup{\AA}}
\begin{document}

\preprint{APS/123-QED}

\title{Annealing for prediction of grand canonical crystal structures:\texorpdfstring{\\}{} Efficient implementation of \texorpdfstring{$n$}{n}-body atomic interactions}

\author{Yannick Couzini\'{e}\orcidlink{0000-0002-5408-8197}}
\email{couzinie.y.aa@m.titech.ac.jp}
\affiliation{
Laboratory for Materials and Structures,
Institute of Innovative Research,
Tokyo Institute of Technology,
Yokohama 226-8503,
Japan
}
\affiliation{
Quemix Inc.,
Taiyo Life Nihombashi Building,
2-11-2,
Nihombashi Chuo-ku, 
Tokyo 103-0027,
Japan
}

\author{Yusuke Nishiya \orcidlink{0000-0001-6526-0936}}
\affiliation{
Laboratory for Materials and Structures,
Institute of Innovative Research,
Tokyo Institute of Technology,
Yokohama 226-8503,
Japan
}

\affiliation{
Quemix Inc.,
Taiyo Life Nihombashi Building,
2-11-2,
Nihombashi Chuo-ku, 
Tokyo 103-0027,
Japan
}

\author{Hirofumi Nishi\orcidlink{0000-0001-5155-6605}}
\affiliation{
Laboratory for Materials and Structures,
Institute of Innovative Research,
Tokyo Institute of Technology,
Yokohama 226-8503,
Japan
}

\affiliation{
Quemix Inc.,
Taiyo Life Nihombashi Building,
2-11-2,
Nihombashi Chuo-ku, 
Tokyo 103-0027,
Japan
}

\author{Taichi Kosugi\orcidlink{0000-0003-3379-3361}}
\affiliation{
Laboratory for Materials and Structures,
Institute of Innovative Research,
Tokyo Institute of Technology,
Yokohama 226-8503,
Japan
}

\affiliation{
Quemix Inc.,
Taiyo Life Nihombashi Building,
2-11-2,
Nihombashi Chuo-ku, 
Tokyo 103-0027,
Japan
}

\author{Hidetoshi Nishimori\orcidlink{0000-0002-0924-1463}}
\affiliation{
International Research Frontiers Initiative,
Tokyo Institute of Technology,
Shibaura, Minato-ku, Tokyo 108-0023,
Japan
}

\affiliation{
Graduate School of Information Sciences,
Tohoku University,
Sendai 980-8579,
Japan
}

\affiliation{
RIKEN, Interdisciplinary Theoretical and Mathematical Sciences (iTHEMS),
Wako,
Saitama 351-0198,
Japan
}

\author{Yu-ichiro Matsushita\orcidlink{0000-0002-9254-5918}}
\affiliation{
Laboratory for Materials and Structures,
Institute of Innovative Research,
Tokyo Institute of Technology,
Yokohama 226-8503,
Japan
}
\affiliation{
Quemix Inc.,
Taiyo Life Nihombashi Building,
2-11-2,
Nihombashi Chuo-ku, 
Tokyo 103-0027,
Japan
}
\affiliation{
Quantum Material and Applications Research Center,
National Institutes for Quantum Science and Technology (QST),
2-12-1, Ookayama, Meguro-ku, Tokyo 152-8552, Japan
}

\date{\today}

\begin{abstract}
We propose an annealing scheme usable on modern Ising machines for crystal
structures prediction (CSP) by taking into account the general $n$-body atomic
interactions, and in particular three-body interactions which are necessary to
simulate covalent bonds.  The crystal structure is represented by discretizing
a unit cell and placing binary variables which express the existence or
non-existence of an atom on every grid point. The resulting quadratic
unconstrained binary optimization (QUBO) or higher-order unconstrained binary
optimization (HUBO) problems implement the CSP problem and is solved using
simulated and quantum annealing. Using the example of Lennard-Jones clusters we
show that it is not necessary to include the target atom number in the
formulation allowing for simultaneous optimization of both the particle density
and the configuration and argue that this is advantageous for use on annealing
machines as it reduces the total amount of interactions. We further provide a
scheme that allows for reduction of higher-order interaction terms that is
inspired by the underlying physics. We show for a covalently bonded monolayer
MoS$_2$ crystal that we can simultaneously optimize for the particle density as
well as the crystal structure using simulated annealing. We also show that we
reproduce ground states of the interatomic potential with high probability that
are not represented on the initial discretization of the unit cell.
\end{abstract}

\maketitle


\section{\label{sec:intro} Introduction}
Crystal structure prediction (CSP) from chemical composition alone is still one of the most difficult problems in materials science, even for the simplest structures~\cite{John}. The reason why this problem is still a challenge is that the variation of possible structures grows exponentially as the number of atoms increases, making an exhaustive search for the most stable structure, i.e.\ finding the global minimum on the Born-Oppenheimer surface, unfeasible even with today's supercomputers. For a small number of atoms, a brute force approach is possible, but reliably finding global optima for larger systems is out of reach of current computers. 

Various approaches to develop searching algorithms that approximate solutions
to the CSP have been developed~\cite{Review}, e.g.\ random
search~\cite{RS1,RS2,RS3,RS4}, simulated annealing (SA)~\cite{wille:1987aa,
wille:2000aa, yin:2022aa}, minima hopping~\cite{MH1,MH2}, evolutionary
algorithm~\cite{EA1,EA2,EA3,EA4} and particle swarm
optimization~\cite{PSO1,PSO2}. Various software suites such as
USPEX~\cite{EA2,EA3,EA4}, CALYPSO~\cite{PSO1,PSO2}, and CrySPY~\cite{CrySPY}
that implement these algorithms continue to be developed and improve upon these
algorithms. However, all of these approaches have one thing in common: as the
system size increases, they become easily trapped by locally stable solutions,
and to escape from these becomes a non-trivial problem. To address this, approaches incorporating experimental data such as X-ray diffraction patterns into the optimization process have been developed~\cite{meredig:2013aa,gao:2017aa,tsujimoto:2018aa}, e.g.\ the data assimilation technique which has been successfully applied to crystal structure and amorphous structure prediction~\cite{adachi:2019aa,yoshikawa:2022aa,zhao:2023aa}.

In recent years, the use of quantum computers has attracted a great deal of attention as a means of searching for globally optimal solutions~\cite{kadowaki:1998aa,durr1999quantum,albash:2018aa,jones:2019aa,mcardle:2019aa,kosugi:2022ab}. 
Quantum computers are characterized by their ability to escape from locally stable solutions and accelerate the search for globally optimal solutions by utilizing the quantum tunneling effect~\cite{denchev:2016aa,albash:2018aa}. Quantum annealing (QA) machines~\cite{kadowaki:1998aa,
hauke:2020aa, king:2022aa, catherine:2019aa,boothby:2020aa} and gate-based quantum computers are the two main current architectures in development. Exhaustive structure search using gate-based quantum computers has been reported recently~\cite{hirai:2022aa,kosugi:2022aa}. In the method described in Ref.~\cite{kosugi:2022aa}, space is divided into meshes, and the presence or absence of atoms on each mesh is represented as a \{0,1\} digital number, allowing the crystal structure to be encoded onto qubits as a bit sequence. On the qubits, various atomic coordination structures can be prepared at once by using the quantum superposition states on the qubits. The idea is to perform exhaustive structural optimization by applying a probabilistic imaginary-time evolution technique reported in~\cite{kosugi:2022ab}.

In this paper, we report a method to perform exhaustive structural optimization
using QA. In particular, we discuss how to reformulate structural optimization as a quadratic unconstrained binary optimization problem (QUBO) or higher-order unconstrained binary
optimization (HUBO). We provide a scheme for implementing an empirical
three-body interatomic potential  on QA hardware, and we provide a detailed analysis of preliminary SA and QA results. In particular we argue that providing more physical information in the form of penalty terms does not necessarily speed up the computation.

The remainder of the paper is structured as followed. In \cref{sec:hubo} we
present the HUBO formulation for the CSP. In \cref{sec:methods} we introduce
the methods and general parameters used for optimization. In \cref{sec:krypton}
we outline the parameters for a Lennard-Jones cluster of Krypton atoms for
which we optimized both structure and density using SA and QA. In
\cref{sec:mos2} we present a covalently bonded MoS$_2$ crystal modeled by a
Stillinger-Weber potential for which we optimized again the structure and
density using SA. We then close with the conclusions in \cref{sec:conclusions}.

\section{HUBO formulation}\label{sec:hubo}
In this section we discuss the construction of our HUBO. In
\cref{sub:csp_encoding} we discuss the notation of our unit cell discretization
and the encoding into a HUBO of the CSP. In \cref{sub:penalty} we disuss the
penalty terms we use and finally in \cref{sub:quadratization}
we discuss a physically-motivated scheme to reduce the interaction terms of
interaction terms of order higher than quadratic.
\subsection{CSP problem encoding and Hamiltonian}\label{sub:csp_encoding}
Consider a unit cell that is spanned by a given basis $\{\vec{a}_i\}$ with
periodic boundary conditions along a chosen set of basis vectors and a set of
 atom species $\mathcal{S}$. We look at a set of
$N$ lattice points $\mathrm{X}$ in this unit cell generated by partitioning each
basis vectors into $g+1$ points and forming the corresponding lattice. The
lattice points have the form $\sum_{i} \frac{k_i}{g} \vec{a}_i$ where $k_i\in \{0,
\ldots, G_i\}$ with $G_i=g$ if we have no periodic boundary conditions along
$\vec{a}_i$ and $G_i=g-1$ otherwise.
Consider a set $b_{x}^{s}$ of binary variables that we define such
that if $b_{x}^{s}=1$ there is an atom of species $s\in \mathcal{S}$ on
$x\in \mathrm{X}$. Assume that we have a set of potential functions $V_m^{s_1,
\ldots, s_m}(x_1, \ldots, x_m)$ for a configuration of atoms of species $s_i$
on $x_i$ for $m\in \{1, \ldots, M\}$. As is usual for interatomic potential
functions we assume that it does not depend on the order in which the argument,
species pairs are supplied, i.e.
\begin{align}
        V_m^{s_1, \ldots, s_m}(x_1, \ldots, x_m)
        \equiv V_m^{s_{\sigma(1)}, \ldots, s_{\sigma(m)}}
        (x_{\sigma(1)}, \ldots, x_{\sigma(m)})
\end{align}
for any permutation $\sigma$. Assuming that we have no periodic
boundary conditions, we define our Hamiltonian as
\begin{align}\label{eqn:hamil}
        \begin{split}
                &H \\& =\sum_{\substack{x\in \mathrm{X}\\ s\in\mathcal{S}}}
                V^{s}_1(x)b_{x}^{s}
                \\
                   &+\frac{1}{2!}
                   \sideset{}{'}\sum_{\substack{x_1,x_2\in\mathrm{X}\\
                           s_1,s_2\in\mathcal{S}}}
                        V^{s_i,s_j}_2(x_1,
                        x_2)b_{x_1}^{s_1}b_{x_2}^{s_2}+\cdots \\
                   &+\frac{1}{M!}
                   \sideset{}{'}\sum_{\substack{x_1,\ldots, x_M\in X\\
                           s_1,\ldots, s_M\in\mathcal{S}}}
                   V_M^{s_1, \ldots, s_M}(x_1,\ldots,x_M)
                   b_{x_1}^{s_1}\cdots b_{x_M}^{s_M},
        \end{split}
\end{align}
where the prime indicates that the $x_i\in X$ should be chosen such that
$x_i\neq x_j$ for any pair $i,j$ (the species are chosen freely). Defined as
such, finding the optimal nuclear structure on the lattice $X$ corresponds to
finding an optimal binary string that minimizes this Hamiltonian, as energy
contributions only arise if all binary variables involved in an interaction are
$1$, i.e.\ all atoms involved in the interaction are present.

Generalising this to the case with periodic boundary conditions requires a
careful consideration of the self-interactions of atoms with their periodic
images and a fitting definition of the Hamiltonian. This is done in detail in
\cref{app:pbc}.

\subsection{Penalty terms}\label{sub:penalty}
\cref{eqn:hamil} allows us to calculate the cohesive energy of a given
configuration (see \cref{app:cohesive}).
Thus, for well-constructed interatomic potentials that accurately model a wide range
of configurations of a material, \cref{eqn:hamil} not only gives the optimal
configuration, but by simultaneously finding the optimal amount of binary
variables that should have the value $1$ we optimize for the optimal density of
atoms in the unit cell.

It is possible to a priori fix a target atom number in the unit cell by adding a
\textit{penalty term} such as
\begin{align}\label{eqn:abs_num_pen}
        P {\left( \sum_{\mathrm{x\in \mathrm{X}}}b_{x}^{s}
        - \mathcal{C}_{s} \right)}^2
\end{align}
to the Hamiltonian for an appropriately large positive $P$ and all
$s\in\mathcal{S}$ where $\mathcal{C}_{s}$ is the target particle number for
species $s$ atoms.
We call this an \textit{absolute penalty term}.

Equivalently, knowing the chemical formula (e.g.
$\mathrm{Al}_2{(\mathrm{SO}_4)}_3$) but not the optimal density, a penalty term like
\begin{align}\label{eqn:rel_num_pen}
        P {\left(
                        \sum_{x\in\mathrm{X}}b_x^{s_1}
                - c_{s_1,s_2}
                        \sum_{x\in\mathrm{X}}b_{x}^{s_2}
        \right)}^2
\end{align}
ensures that the ratios of atoms are respected, where $c_{s_1,s_2}$ is the
target ratio (in the above example $c_{{\rm S},{\rm O}}=1/4$). This penalty
term allows for finding the optimal density in the range that the ratio is
respected. We call this a \textit{relative penalty term}.

\subsection{Reduction of interaction terms}\label{sub:quadratization}
Interatomic potentials will usually include a cutoff distance. To reduce
pairwise interaction terms, it is crucial to choose the right penalty terms as
an absolute penalty term will introduce interactions between any pair of binary
variables for the same species, even if their pairwise distance is higher than
the cutoff distance. Similarly, relative penalty terms introduce pairwise
interactions between any pair of binary variables of the two involved species.
Choosing the wrong penalty terms can make the difference between having a
sparse or fully connected graph of pairwise interactions. Ideally, no penalty
terms would be introduced, but this is dependent on the quality of the chosen
potential.

The number of interaction terms for the cubic or higher order terms in the HUBO
will be orders of magnitudes higher than for the pair interactions. Often,
alongside the total number of spins, the density of the interaction graph is
the main bottleneck for modern annealing machines~\cite{kanao:2022aa,
boothby:2020aa} and as such it is crucial to devise schemes that reduce the
interaction number beyond just applying a cutoff. To this end we use the
`deduc-reduc' method from~\cite{tanburn:2015aa}. In particular, we make the
assumption that if the pairwise interaction between two binary variables is
higher than a user-set threshold $T$, then any higher-order interaction containing
this pair can safely be set to $0$ without influencing the ground state.
At the same time we replace any pairwise interaction $J_{ij}$ by $\min(J_{ij},
T)$. The
intuition behind this is that for the interatomic potentials we use in this
work, the pairwise interaction rapidly increases if the atoms are too close,
and thus the ground state does not contain atoms on the two involved locations
and we do not need to evaluate the higher-order terms. This is a simplification
that does not loose any generality with respect to the ground state of the HUBO
and which in particular also does not require any a priori knowledge like
atomic radii of the involved species.

\section{Methods}\label{sec:methods}
We will find optimal binary strings for the HUBO problems using SA and QA. In
this section we outline the notation, parameters and settings we used for the
optimization.
\subsection{Simulated Annealing}\label{sub:sim_ann}
Simulated annealing is a classic algorithm for optimizing cost functions with
several local minima~\cite{bertsimas:1993aa}. We assume some basic knowledge of
the algorithm and will only discuss the specifics of our implementation. We use
a geometric cooling schedule
\begin{align}
        T(x) = T_{\max} {\left(
        \frac{T_{\min}}{T_{\max}}\right)}^{x/N_{\mathrm{steps}}}, \quad x\in
        [0, N_{\mathrm{steps}}],
\end{align}
where $T_{\min}$ and $T_{\max}$ are the minimum and maximum temperature. The
number of steps $N_{\mathrm{steps}}$ is the number of Monte Carlo steps per
spin to perform.

Choosing the right neighbourhood for a configuration in SA
(i.e.\ defining legal transitions of the Markov chain) is crucial and generally
one aims to have a smooth energy landscape with not too rugged local
minima~\cite{solla:1986aa, eglese:1990aa, henderson:2003aa}. Traditionally,
SA for HUBOs performs single bit flips. As this is equivalent
to removing or adding an atom from the configuration, especially in the
presence of penalty terms, this can be a costly operation. Thus, for each step
in the schedule we loop over every binary variable and attempt to flip it and
then we loop over every opposite valued pair in the current configuration and
attempt to exchange their values. This latter flip moves an existing atom to a
random location and does not break penalty terms such as the absolute
penalty~\eqref{eqn:abs_num_pen} or relative penalty~\eqref{eqn:rel_num_pen},
thus ensuring a smoother energy landscape. So when we speak of Monte Carlo
steps per spin we mean that we attempt $N\cdot |S| + \binom{N\cdot
|\mathcal{S}|}{2}$ spin flips where $N\cdot |\mathcal{S}|$ is the (unreduced) binary variable number.

\subsection{Quantum Annealing}\label{sub:quant_ann}
We also assume familiarity with the basic concepts
of quantum annealing~\cite{kadowaki:1998aa,albash:2018aa}. We use the Advantage
system available through the D-Wave leap cloud service~\cite{mcgeoch:2022aa}.
Our HUBO and QUBO problems are very densely connected and if the cutoff of the
potential function is large enough or the system small enough, the problem
might even be fully connected. Embedding these onto the Pegasus architecture of
the Advantage system~\cite{boothby:2020aa} requires us to calculate a minor
embedding~\cite{choi:2008aa, choi:2011aa, zbinden:2020aa}. Instead of manually
calculating an embedding best fit for our problem, we use the standard
implementation for clique embedding in the D-Wave Ocean SDK.\@ This procedure
can lead to results with broken chains which require a fitting unembedding.
While there is evidence that designing a problem specific unembedding
algorithm~\cite{pelofske:2020aa} can be advantageous we choose the simple majority vote which sets
the binary value of a chain to the one that occurs most often on the chain.

\subsection{Benchmarking}%
\label{sub:tts}
For benchmarking the various optimization schemes for the HUBO and QUBO
formulation we use the time-to-solution~\cite{prielinger:2021aa,
kadowaki:2023aa} given by
\begin{align}
        \mathrm{TTS}(\tau) = \tau
        \frac{\ln(1-p_r)}{\ln(1-\mathbb{P}_{\rm GS}(\tau))}=
        \tau \frac{\ln(0.01)}{\ln(1-\mathbb{P}_{\rm GS}(\tau))},
\end{align}
where $\tau$ is the running annealing time as measured on the local machine and
$\mathbb{P}_{\rm GS}(\tau)$ is the probability of the corresponding algorithm to
return the ground state with a running time of $\tau$. The time-to-solution can
be understood as the average time it takes to get the ground state with
probability $p_r$ which we set to $0.99$.

\section{Krypton system}\label{sec:krypton}
In this section we introduce an LJ cluster system consisting of Krypton atoms
in \cref{sub:lj_setup} and the related SA and QA results in
\cref{sub:lj_results}.
\subsection{Setup}\label{sub:lj_setup}
\begin{figure}
        \includegraphics[scale=0.125]{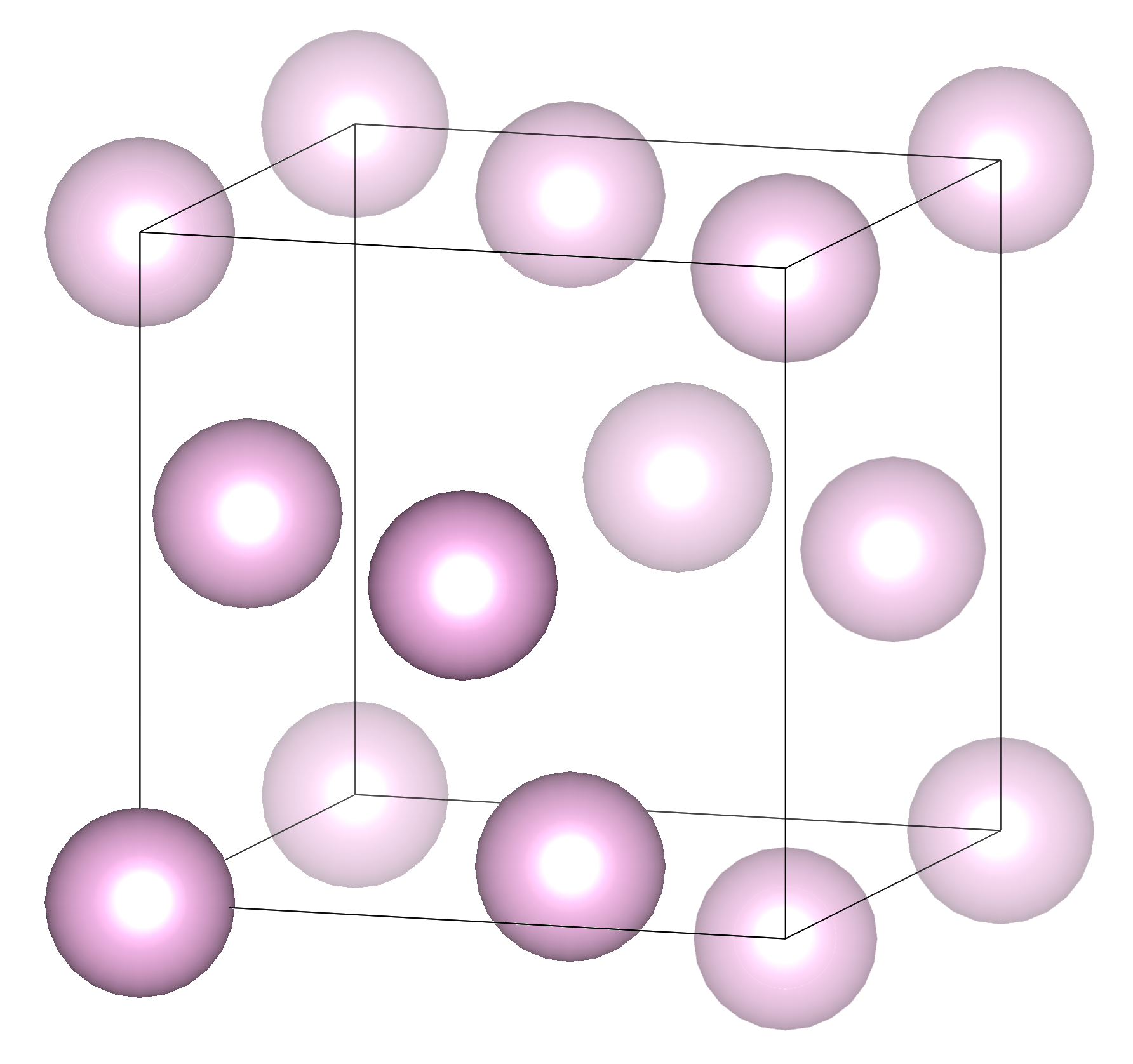}
        \caption{The target FCC configuration of the Krypton
        system with Krypton atoms in pink (graphics due to Vesta). The solid
        atoms on the origin and the three incident face centers are the
        locations encoded in the HUBO while the remaining transparent ones are
        copies due to the periodic boundary conditions and not part of $\mathrm{X}$.}\label{fig:lj_gs}
\end{figure}

For the calculation of the potential functions we rely on the Open
Knowledgebase of Interatomic Models (OpenKIM)~\cite{tadmor:2011aa}. In
particular we will look at a three dimensional cubic unit cell of side length
$5.653$\angstrom{} with the Lennard-Jones potential parameters due to Bernades
for Krypton~\cite{bernardes:1958aa, elliott:2011aa, tadmor:2011aa,
tadmor:2020aa, tadmor:2020ab} and periodic boundary conditions along all three
basis vectors. We will look for the ground state configuration of Krypton atoms
in this unit cell discretized into a equipartitioned lattice of size $g^3$,
which is equal to the face-centered cubic configuration and can be seen in
\cref{fig:lj_gs}. The energy of the FCC configuration is $-0.431$eV and for any
interaction value $J_{ij}$ we take $\min(J_{ij}, 1\text{eV})$. While for the
SA calculations this is not strictly necessary, it helps for
the QA calculations as the energy range is normalised to be
between $0$ and $1$ on D-Wave machines, thus upper bounding the energy ensures
that the physically interesting energy range takes up a larger portion of the
renormalised energy range. We will simply refer to this system as the Krypton
system. We perform SA calculations without any penalty terms and with an
absolute number penalty term setting $\mathcal{C}_{\rm Kr}=4$, we call the former
grand canonical and the latter microcanonical. As the unit cell is smaller than
the cutoff distance of the potential, even the grand canonical calculation QUBO
is fully connected. We use a penalty strength of $P=1$, and vary the
temperature from $10^{-2}$ to $10^{-4}$. The various probabilities correspond
to the measured probability
across $1000$ annealing runs.

 Since the systems are
fully connected, for the QA calculations, we simplify the QUBO by fixing the binary variable for the
origin to be $1$ and removing any binary variable that had an interaction with
the origin of more then $1$eV. This corresponds in essence to removing the
translational invariance of the problem.
Further, we use
pausing~\cite{chen:2020aa,gonzalez-izquierdo:2022aa}. We use a base length of
the schedule of $20\mu$s and we pause for $3\mu$s. We consider the success
probability, i.e.\ the ratio of obtained ground states over $40000$ annealing
runs, plotted against the pause location $s_p\in (0, 1)$ so that the
dimensionless time in the annealing schedule goes from $0$ to $s_p$ at
$(17\cdot s_p)\mu$s until $(17\cdot s_p)\mu\text{s} + 3\mu$s and then goes to $1$
linearly until $20\mu$s. We use a chain strength of $1.28$. These parameters
were heuristically found to provide reasonable results.

\subsection{Results and discussions}\label{sub:lj_results}
\begin{figure}
        \centering
        \includegraphics[width=0.48 \textwidth]{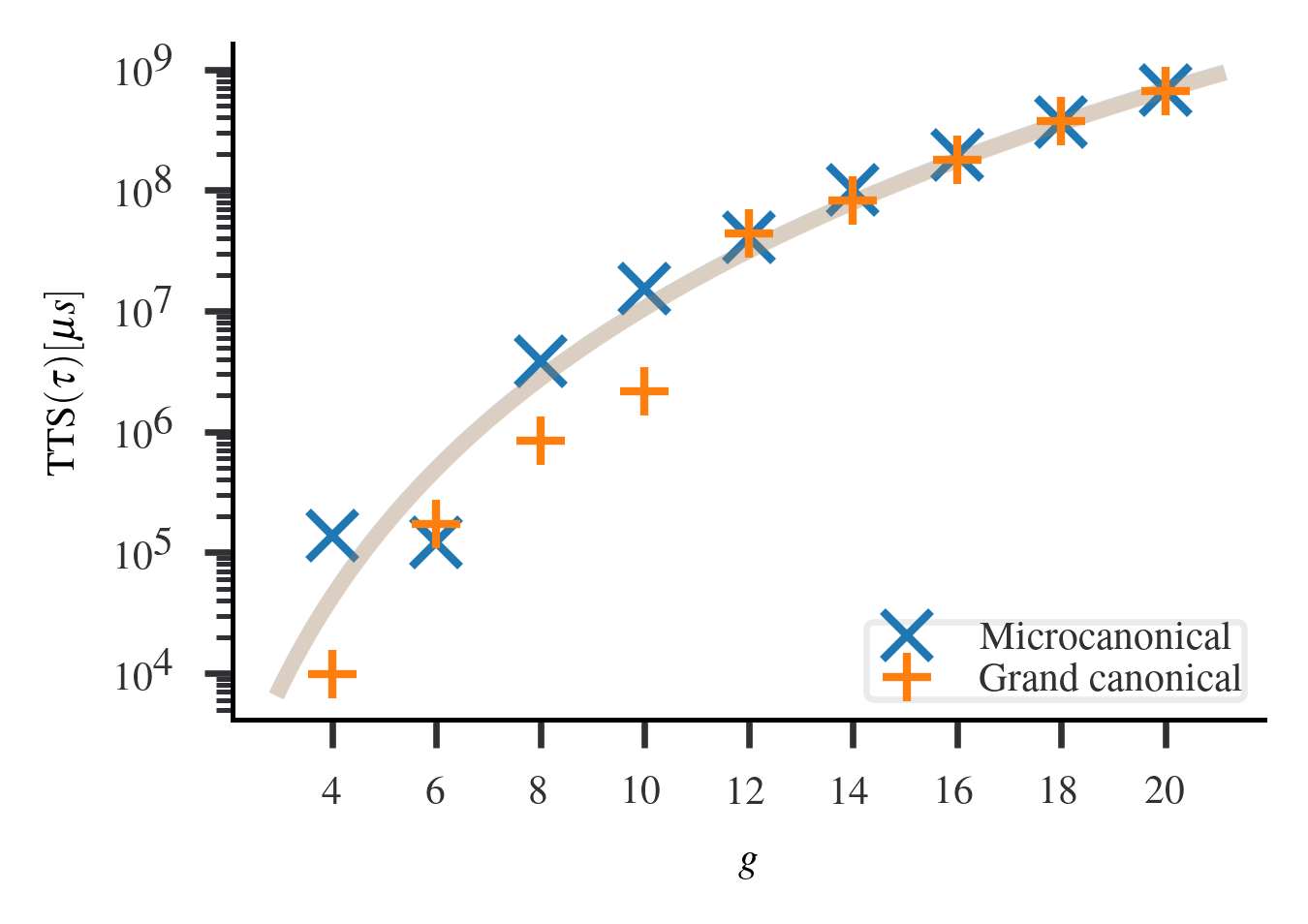}
        \caption{The SA time to solution results for the Krypton system with a penalty
        term in blue crosses and without in orange plus-symbols plotted against various grid
        granularities $g$. The solid line corresponds to a fit
        of the measured points to $a (N
        +\binom{N}{2})$
        where $a=21.015$ is the fitting parameter.}\label{fig:lj_results}
\end{figure}
In \cref{fig:lj_results} we plot the TTS against various grid
spacings $g$ for SA calculations for the grand and microcanonical system. We
performed SA until we found the ground state FCC configuration with a
probability of more than $90\%$ and take the minimum TTS across the schedule
steps as the data point for $g$. This takes at most $30$ schedule steps for
both systems and it is apparent that both systems have comparable performance.
In particular note the fit to the function $N+\binom{N}{2}$, which is the
scaling of the number of flips the SA algorithm attempts with
the spin number $N$. There are two main mechanisms that increase the required
TTS. The first is that, as we attempt more spin flips per schedule step with
increasing $g$, SA requires more time per
schedule step to perform the increasing amount of flips. The second is that
with increasing $g$ the atoms have more fine-grained displacement possibilities
so that there are more local minima of the QUBO problem with energies closer to
the actual ground state leading to an increased time to escape the local minima
to find the ground state.

If the global minimum were harder to find due to increasing amounts of
local minima, we would expect an increasing number of required schedule steps
with increasing $g$. What we see is that the fit $a(N+\binom{N}{2})$ with a
constant $a=21.015$ reconstructs the data well for $g\ge 12$ for both systems.
Thus there is no significant scaling $\sim TTS(\tau)/(N+\binom{N}{2})$ of the
required scheduled steps with $g$ for the microcanonical and the grand
canonical system.

\begin{figure}
        \centering
        \includegraphics[width=0.48 \textwidth]{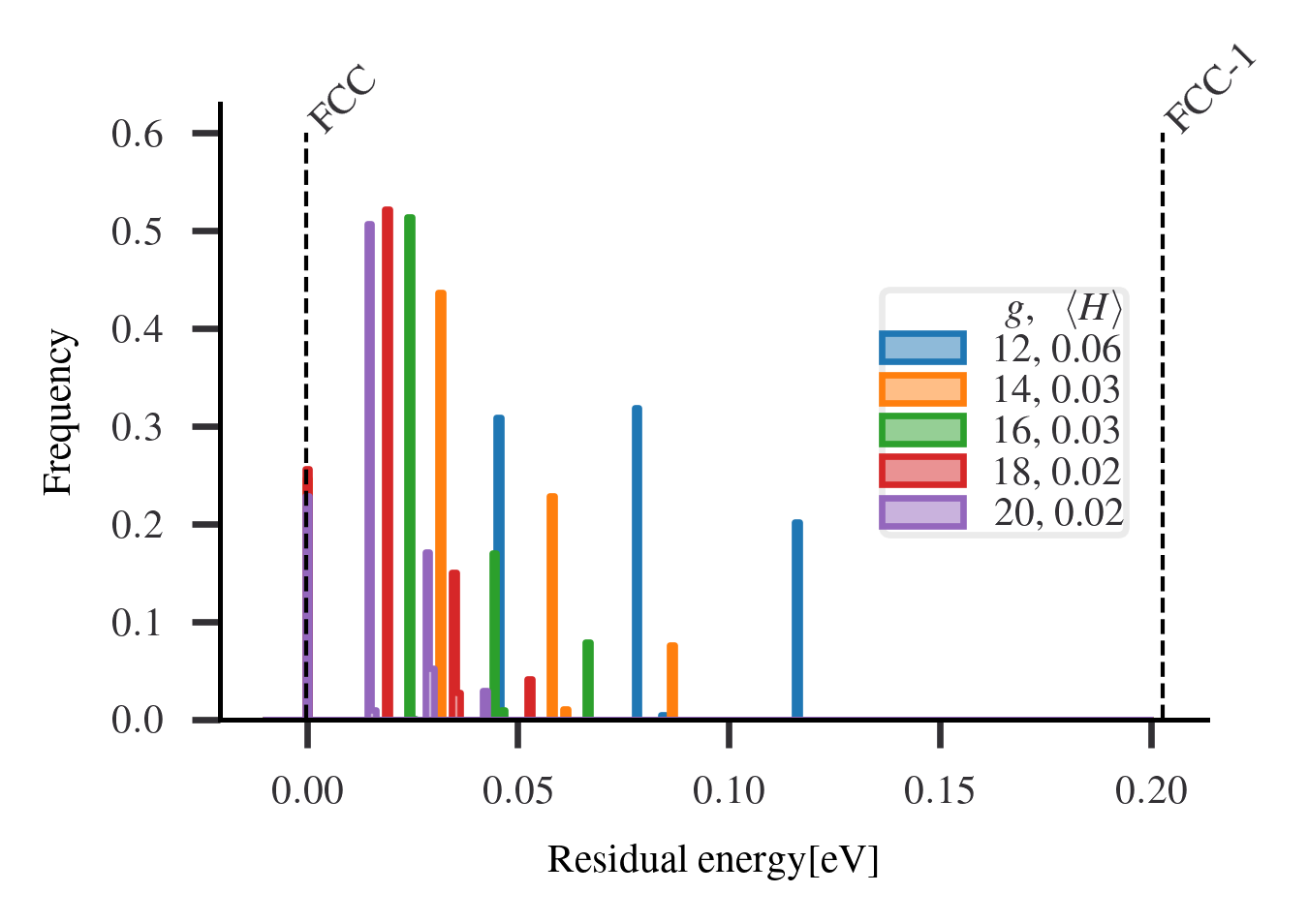}
        \caption{The histogram for the residual energy of the Krypton system after
                running SA for $3$ Monte Carlo steps per spin and the
                various $g$ together with their average residual energy, i.e.\ energy
                above the ground state,  $\langle H\rangle$. This is the full
                histogram, no results have been cut.}\label{fig:lj_histogram}
\end{figure}
Further, in \cref{fig:lj_histogram} we show a representative energy histogram
for the grand canonical calculations with $3$ Monte Carlo steps per spin for $g\in \{12, 14,
16, 18, 20\}$. Despite not putting any particle number restrictions
the annealing process, even for this low amount of schedule steps, only
returns solutions with the correct atom density and in fact all returned
energies are
lower than the first excited state energy
corresponding to an FCC configuration with an atom taken out (see
\cref{fig:lj_loc_mins} in the appendix), a state we call FCC$-1$. Using the
Broyden–Fletcher–Goldfarb–Shanno (BFGS)
algorithm~\cite{broyden:1970aa,fletcher:1970aa,goldfarb:1970aa,shanno:1970aa}
to converge to a local minimum off the grid $X$
we confirmed that all states with $4$ atoms converge to the ground state meaning
that the TTS of the combination of annealing combined with BFGS is
considerably lower than that of only annealing.
\begin{figure}
        \centering
        \includegraphics[width=0.48 \textwidth]{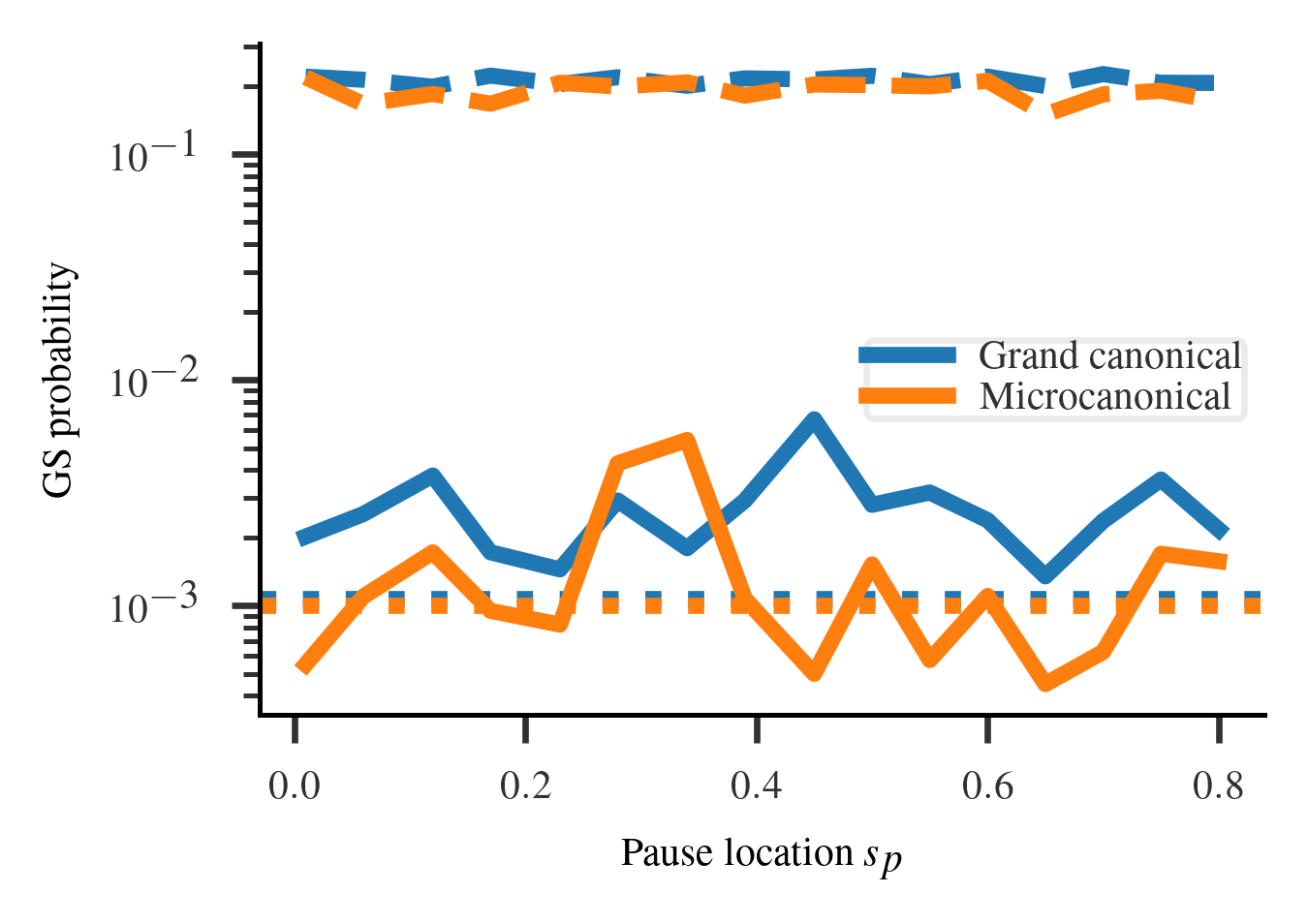}
        \caption{Ground state probabilities for the $g=4$ Krypton system using
                the D-Wave Advantage 4.1 system with various pause locations
                $s_p$ ranging from $0.01$ to $0.8$. In blue the grand canonical
                calculation and in orange the microcanonical with a penalty
                strength of $0.05$. The dashed lines correspond to the ground
                state probability after applying BFGS on the results from the solid
                lines and the dotted line to the probability of running annealing with no
                pauses and an annealing time of $18.9\mu$s.}\label{fig:pause_locs}
\end{figure}

We also confirmed these tendencies on the D-Wave Advantage 4.1 system available
on D-Wave Leap. We performed calculations only for the $g=4$ system since the
minor embedding for the $g=6$ system had chain lengths of up to $20$ spins
which proved too hard to optimize. In \cref{fig:pause_locs} we plot the pause
location $s_p$ against the success probability for the grand and
microcanonical system for just QA with pausing
and without pausing and a schedule length of $18.9\mu$s and quantum
annealing with pausing followed by BFGS. The penalty strength in the
microcanonical calculations is $0.05$ as it provided the best ground state
probability. First we see that pausing improves the performance as for both
systems the probabilities without pausing are around $0.001$ and with pausing
the maximum probabilities for the grand canonical system are $0.0067$ at
$s_p=0.45$ and $0.005425$ at $s_p=0.34$ for the microcanonical one. Since there
are no same-density local minima, performing BFGS optimization on the results
with pausing, is equivalent to looking at the results that have the correct
density. We see that for QA+BFGS calculations both systems have success
probabilities between $0.15$ and $0.22$ with the grand canonical consistently
having a higher probability.

Without pausing QA has a TTS of
around $0.9*10^5\mu$s comparable with the microcanonical system TTS for SA in the
$g=4$ case (see \cref{fig:lj_results}). With pausing we find a TTS of $13700\mu$s and $16931\mu$s
respectively for the grand and microcanonical system providing comparable
times to the grand canonical SA calculations albeit the QA calculations are a
bit slower. Thus we find no indications of a quantum speedup. Possible reasons
for this result may include the embedding of full connectivity on the sparse
hardware graph and noise effects. We leave it for future research to analyse
this problem with a wider set of parameters and using more intricate embedding
techniques.

Note though, that while the SA+BFGS algorithm did not provide any other minima than the
global one, QA+BFGS returns the FCC-1 configuration with probabilities
between $0.3$ and $0.33$ across all pause locations $s_p$ for the grand
canonical system and $0.21$ and $0.25$ for the microcanonical system. Thus
while we might not expect a quantum speedup there might be an advantage due to
the higher breadth of results returned by QA compared to SA allowing a
wider exploration of the potential energy landscape.

Summarizing, we see that also for QA, at least in this very simple system,
there are no performance costs in leaving out the penalty and in fact we can
expect performance increases confirming the tendencies found in SA.

\section{\texorpdfstring{MoS$_2$}{MoS2} system}\label{sec:mos2}
In this section we introduce a MoS$_2$ system governed by the three-body
Stillinger-Weber potential in \cref{sub:mos2_setup} and the related SA results in
\cref{sub:mos2_results}.
\subsection{Setup}\label{sub:mos2_setup}
\begin{figure}
        \includegraphics[scale=0.125]{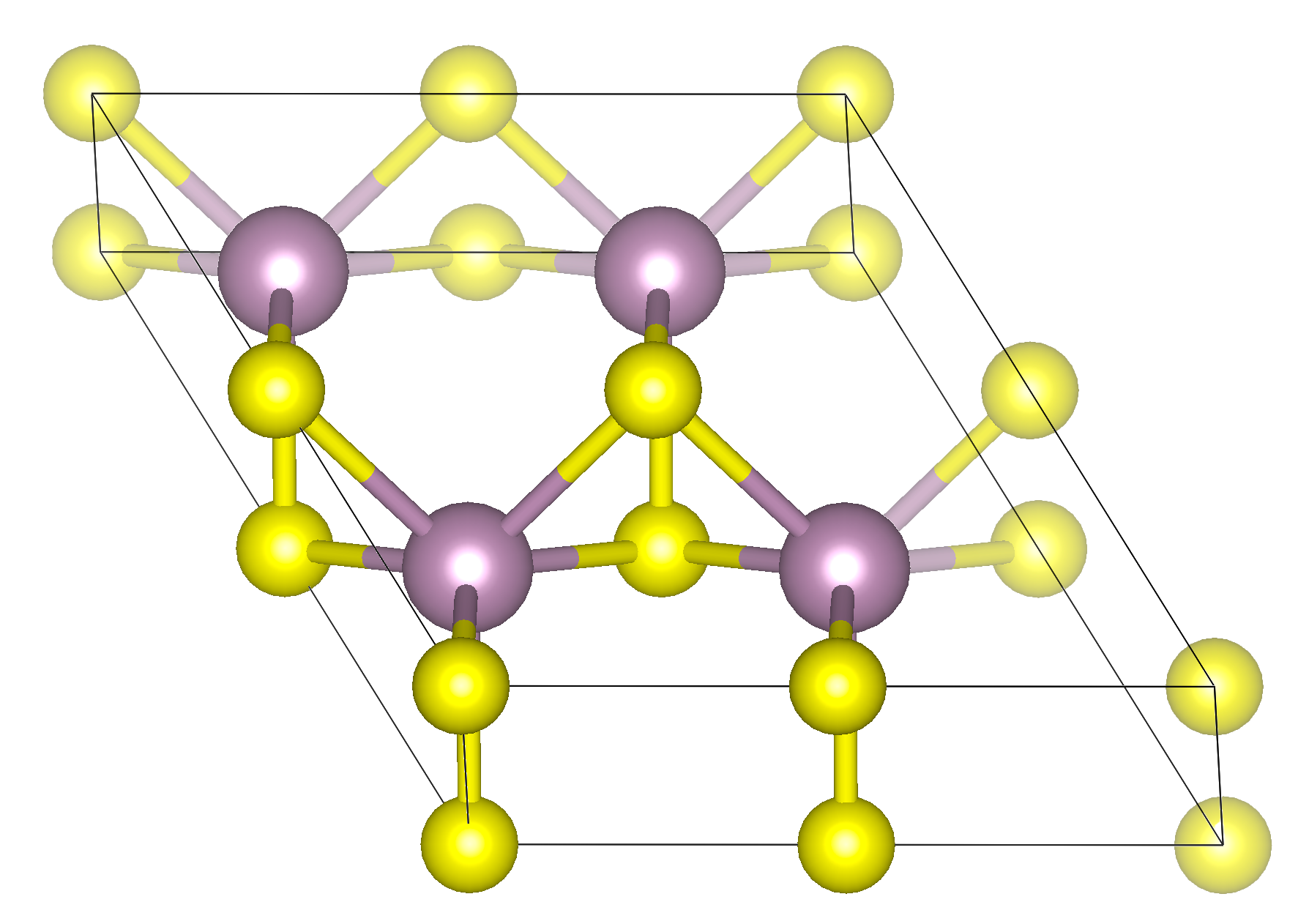}
        \caption{The target $2\mathrm{H}$ ground state configuration of the MoS$_2$
                system with Sulfur in yellow and Molybden in violet (graphics
                due to Vesta). The bottom six sulfur atoms on the boundary, two
                at $(\vec{a}_1+\vec{a}_2)/2$ and four Molybden atoms with
                $z$-coordinate given by $\vec{a}_3/2$ are the locations encoded
                in the HUBO (non transparent atoms). The remaining ten Sulfur atoms
                (transparent) are copies due to the periodic boundary conditions and not part
                of $X$.}\label{fig:mos2_gs}
\end{figure}
For the second system we consider the Stillinger Weber
potential~\cite{stillinger:1985aa, stillinger:1986aa} which is a simple
three-body potential that reflects covalent bond dynamics. We use the
parametrization for hexagonal monolayer Molybden-Disulfide due to Wen et
al.\cite{wen:2017aa,wen:2018aa,kurniawan:2022aa,kurniawan:2022ab}. We do this
on the supercell consisting of a $2\times 2$ lattice of hexagonal lattice unit
cells with a single unit cell having a lattice constant of $3.20$\angstrom{} and thickness of
$3.19$\angstrom{}. Thus the lattice vectors for our system are $\vec{a}_1 =
(3.2\angstrom,
-\sqrt{3}\cdot3.2\angstrom,0)$, $\vec{a}_2= (3.2\angstrom,
-\sqrt{3}\cdot3.2\angstrom,0)$, $\vec{a}_3=(0, 0, 3.19\angstrom)$ We build the
lattice by partitioning both $\vec{a}_1$ and $\vec{a}_2$ into $g=6$ equal parts
each and applying periodic boundary conditions and partioining $\vec{a}_3$ into
three equal parts without periodic boundary conditions. Thus the amount of
required bits scales like $6g^2$, where the additional $2$ comes from the
amount of species. The target ground state is the $2\mathrm{H}$
configuration (see \cref{fig:mos2_gs}) and has an energy of $-55.5283$eV. The first
excited state that we expect to see is the $1\mathrm{T}$ configuration, with the same amount of atoms and an energy
that is $1.4755$eV above the ground state (see~\cref{fig:mos2_loc_mins} in the
appendix). We will
refer to this system as the MoS$_2$ system.

We use our deduc-reduc with a threshold of $10$eV which in this particular case
reduced the amount of non-zero three-body interaction terms by $18.8\%$ (from
$1573728$ to $1277267$) in the $g=6$ system. Any lower threshold seemed to impact the ground state
configuration on our SA calculations. There is no general-use
scheme known to the authors, that would allow to quadratize this HUBO so as to
make it runnable on any modern Ising
machine~\cite{boros:2014aa, anthony:2017aa, dattani:2019aa} and so while our
deduc-reduc step reduces the interactions it can only be a first step in
conjunction with other approaches yet to be found and we perform no QA for this
system.

We perform SA for the system with both absolute penalty terms
($\mathcal{C}_{\mathrm{Mo}}=4$, $\mathcal{C}_{\mathrm{S}}=8$) and relative penalty terms
($\mathcal{C}_{\mathrm{Mo}, \mathrm{S}}=1/2$). For simplicity we call the former the absolute
system and the latter the relative system. Grand canonical calculations as in
the Krypton system without penalty terms do not work for this potential, as it
is more favourable to produce configurations with a single atom species rather
than a MoS$_2$ mix, so we limit our analysis to the relative and absolute
system and recall that the former retains the function of simultaneously
optimizing for the atom density. The number of pairwise interaction
terms without interactions increases by $1.2\%$ using the absolute penalty
(from $21420$ to $21708$) and by $8.4\%$ using the relative penalty ($23220$),
underlining again the importance of finding potentials that can be used without
penalties to reduce the number of pairwise interactions necessary. In fact,
since this potential is parametrized for hexagonal MoS$_2$ we cannot expect it
to yield accurate results for non-hexagonal configurations. This is a problem
that does not pertain to the parametrization but the Stillinger-Weber potential
in general. Since this one of the simplest three-body potentials we use it
anyway for this proof-of-concept calculation.

We use a penalty strength of $P=10$ and a temperature range of $10$ to $0.1$
for SA. The various probabilities correspond to the
measured probability across $1000$ annealing runs.

\subsection{Results and discussions}\label{sub:mos2_results}
\begin{figure}
        \centering
        \includegraphics[width=0.48 \textwidth]{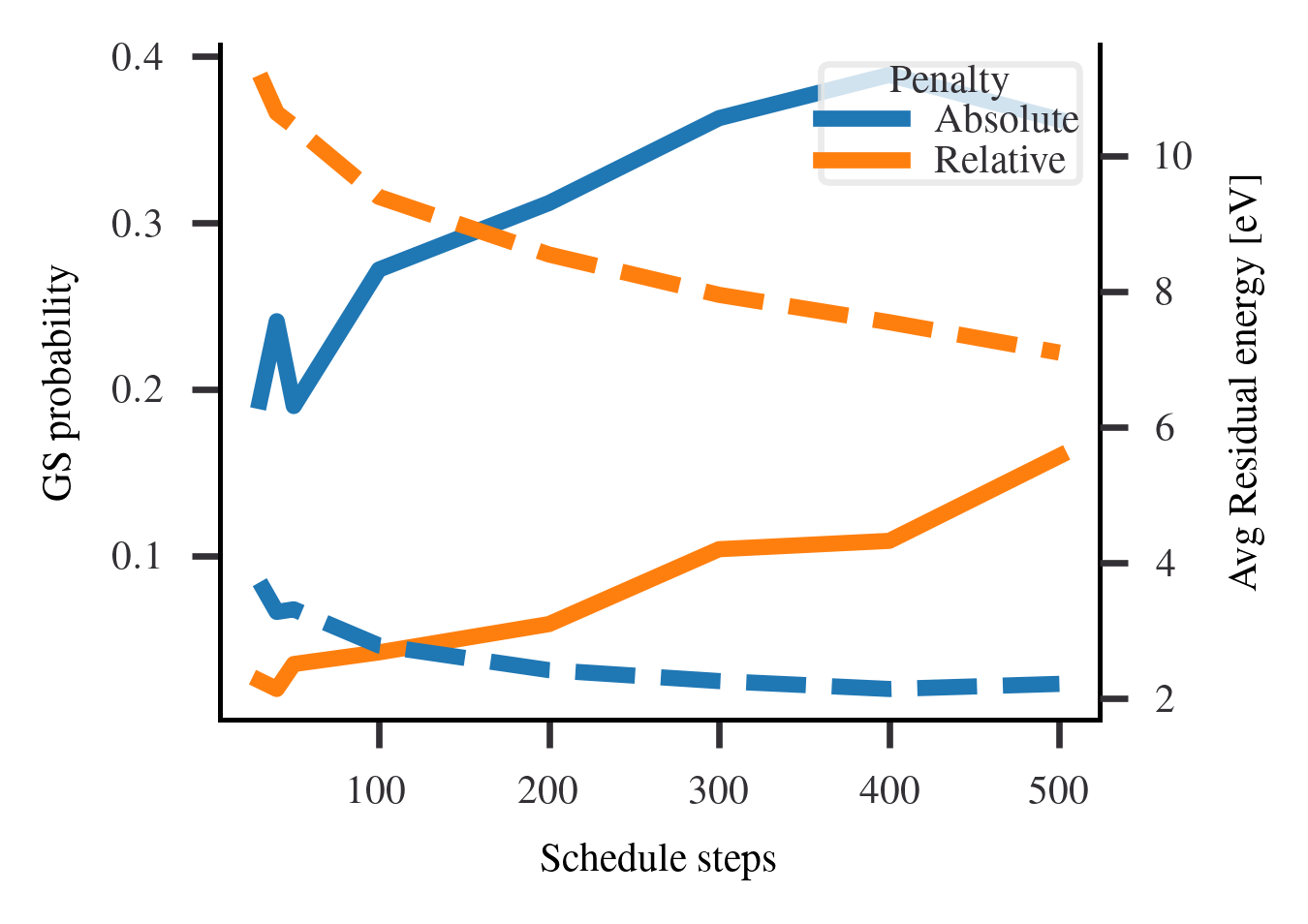}
        \caption{Plot in solid lines of the ground state probability for SA for the
                MoS$_2$ system with schedule steps going
                from $2$ to $500$ for both relative penalties and absolute penalties
                (blue and orange respectively). The scale for the probability is to the
                left. In dashed lines the average residual energy with the
                corresponding scale to the right.}\label{fig:mos2_gs_prob}
\end{figure}
The MoS$_2$ system proves harder to optimize than the Krypton system. In
\cref{fig:mos2_gs_prob} the ground state probabilities for schedule steps going
from $2$ to $500$ are plotted. As opposed to the Krypton system where even for
$g=20$ we need only $30$ schedule steps to reach a ground state probability of
above $0.9$ we see that it hovers around $0.4$ for the absolute penalty and
around $0.15$ for the relative penalty at $500$ schedule steps. In particular
note that here the used penalty terms have an effect on the ground state
probability, and that supplying more information (in form of the absolute
penalty) leads to higher ground state probabilities. As expected the ground
state probability increases with increasing amount of schedule steps but the
slope does not offset the increase in calculation length and so the TTS turns
out to be minimized for a number of schedule steps in the single digits for
both system. In \cref{fig:mos2_gs_prob} the average residual energies are
plotted and we see that both systems seem to converge to an average residual
energy that is well above the target $0$eV.
\begin{figure}
        \centering
        \includegraphics[width=0.48 \textwidth]{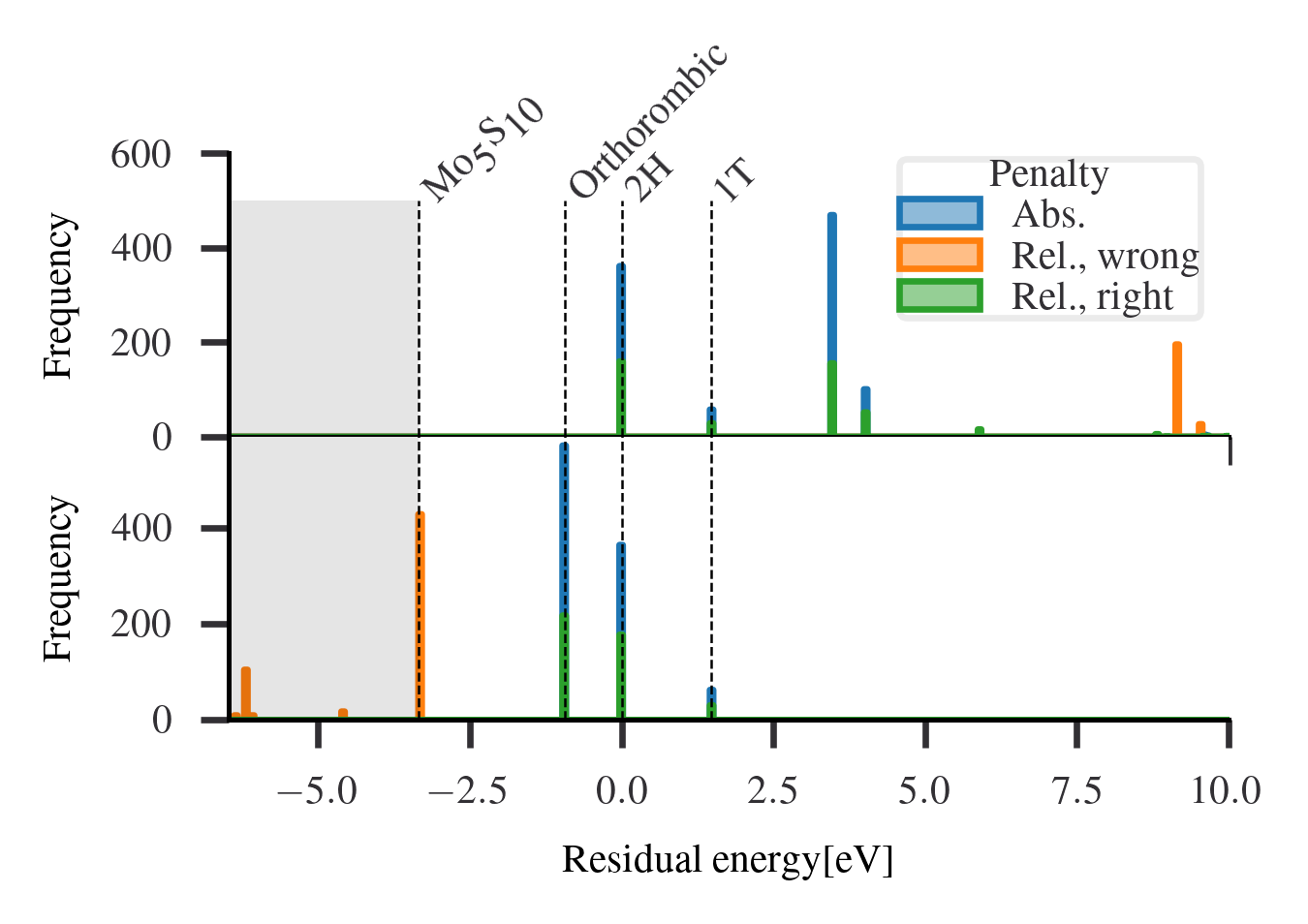}
        \caption{Histogram of the residual energy for the MoS$_2$ system with
                SA with $500$ schedule steps (top) and
                SA + BFGS (bottom) applied to the MoS$_2$
                system with the absolute number penalty \cref{eqn:abs_num_pen}
                in blue and the relative number penalty \cref{eqn:rel_num_pen}
                in orange for results with sub-optimal density and green for the
                optimal density. Found local minima are marked by a dotted line and the shaded
                area to the left (see \cref{app:loc_mins} for the
                configurations). This is not the full histogram, i.e.\ there
                are configurations with energies higher than $10$eV.}\label{fig:mos2_histogram}
\end{figure}
To understand this, consider the energy histogram in \cref{fig:mos2_histogram}
for the resulting states of only SA (top) and SA followed by BFGS with the same potential (bottom) after $500$
schedule steps. First, note that despite not fixing an absolute number of atoms
in the relative penalty, we find the correct density of Mo$_4$S$_8$ in $42.8\%$
of the configurations (in green in \cref{fig:mos2_histogram}) and that the
average residual energy for the states with the correct density is $2.3826$eV
while it is $10.6117$eV for the states with the wrong density (in orange) so
that the relative penalty calculations allow for simultaneous optimization of
the atom density and the optimal configuration. The probability to obtain
either $2\mathrm{H}$ or $1\mathrm{T}$ configurations is $42\%$ for the absolute penalty system
and $18.9\%$ for the relative penalty system. To understand the physical nature
of the remaining local minima, which form the majority of found states, we
performed BFGS on all the resulting states from SA. While the
probability for $2\mathrm{H}$ and $1\mathrm{T}$ rose to $42.8\%$ and $20.9\%$ for the absolute
penalty and relative penalty system respectively we see that most states
converge to a local minimum that has an energy below that of $2\mathrm{H}$. First, for
the relative penalty system we see that $57.2\%$ of all observed configurations
have $5$ Molybden atoms and $10$ Sulfur and form configurations that have an
energy that is more than $2.5$eV lower than that of $2\mathrm{H}$. In
\cref{fig:mos2_histogram} we only shade the region as the BFGS algorithm does
not converge well for these configurations so that we do not get well formed
peaks but rather a distribution in the shaded area. The next lower state is a
state we call orthorombic (see \cref{app:loc_mins} for an image of both the
orthorombic and an example Mo$_5$S$_{10}$ configuration) and has an energy that
is $0.9313$eV lower than that of $2\mathrm{H}$. We find this configuration with a
probability of $21.8\%$ for the relative penalty system and $57.2\%$ for the
absolute penalty system.

Using the Vienna ab initio simulation
package~\cite{kresse:1993aa,kresse:1996aa,kresse:1996ab} with the projector
augmented-wave method~\cite{blochl:1994aa,kresse:1999aa} we find that
the energy of the $2\mathrm{H}$ configuration is in fact the lowest of the four found
local minima, followed by the $1\mathrm{T}$, the orthorombic and finally the
Mo$_5$S$_{10}$ configurations. The fact that this order is not represented is due
to the fact that the Stillinger-Weber potential is parametrized to model
hexagonally ordered MoS$_2$ configurations and thus does not correctly model
other configurations. The potential is not fit to provide new physical
insights in our application and these results should be taken merely as a proof of concept.

Noteworthy about these results is that, despite the orthorombic and locally
optimal Mo$_5$S$_{10}$ states not being representable on the discretization of
the unit cell, the combination of SA and BFGS managed to find
these states in a majority of attempts. This is a strong indication that if we
are able to provide a fitting potential or directly a fitting HUBO we can find
a wide array of globally and locally optimal configurations even if they are
not part of the initial discretization. Thus, in particular it might suffice to
have rougher discretizations with spin numbers that fit onto current quantum
hardware instead of trying to be fine grained enough to represent all possible
local minima.

\section{Conclusions}\label{sec:conclusions}
In this paper we have presented an annealing scheme for crystal structure
prediction based on $n$-body atomic interactions. We discretized a given unit
cell with a lattice and placed binary variables on the lattice points to
express the existence or non-existence of an atom at every grid point. In
particular this is done for $3$-body atomic interactions which is the minimum
order necessary for covalent crystals. We solved the resulting HUBOs using
SA and QA giving insights into the crystal structure. We
have shown that a grand canonical calculation without penalty terms allows for
the simultaneous optimization of both the nuclear structure as well as the
particle density inside the unit cell. Further, we have also shown evidence
that the difficulty of solving the nuclear structure problem does not
necessarily scale with the mesh size. These results show that it might not
always be advantageous to put all the available information into the QUBO to
speed up calculations in particular as this also increases the amount of total
interaction terms the reduction of which is crucial for embedding problems into
modern hardware with limited graphs.

We also considered a Molybden-Disulfide monolayer system modeled by a
three-body Stillinger-Weber type potential. Using our interaction number
reduction scheme we reduced the amount of cubic interactions by $18.8\%$ while
maintaining physical accuracy to the extent of the used potential. We have
shown that the potential contained unphysical ground states that are due to the
limited transferability of the potential outside the context of hexagonal
monolayer MoS$_2$. While these results do not provide physical insights, we
show that our algorithm reproduces the ground state of the system even if they
are not representable on the chosen discretization of the unit cell in the
annealing step of the algorithm. Thus, while we could only optimize the
roughest discretization for the Krypton system on the D-Wave quantum annealer, this
could be a hint that rougher discretizations, that are easier to embed onto
quantum annealers, are enough for the local optimization algorithm to find a
wide array of ground state and locally optimal configurations.

An immediate future research question is to choose a more fitting potential to
construct a HUBO that accurately models a wide array of covalent crystal
configurations to test the performance with rough unit cell meshes on larger
unit cells.

Another research direction is to investigate the nature of returned local
minima by QA and to confirm the tendency we found where QA provided a more
varied insight into the energy than SA which tended to favour only ground
states.

\textit{Note added.} During the writing of this manuscript we have become aware
of a similar proposal for the construction of the QUBO~\cite{gusev:2023aa} for
ionic crystals.
That paper does not address higher-order optimization problems and thus does
not address covalent bonds and did not consider the grand canonical case, their
focus is on classical computation and providing guarantees that ground truths
to the crystal structure prediction problem
are found using their algorithm. They have similar findings with respect to the
reproducibility of the ground state even if it is not contained in the initial
discretization.\@

\begin{acknowledgments}
The authors wish to thank Shu Tanaka, Yuya Seki, Ryo Tamura for the insightful
discussions at the draft stage of this paper and Jun-ichi Iwata for the
discussions concerning the HUBO creation. This work was supported by JSPS
KAKENHI as ``Grant-in-Aid for Scientific Research(A)'' Grant Number 21H04553.
The computation in this work has been done using the TSUBAME3.0 supercomputer
provided by the Tokyo Institute of Technology. The work of H. Nishimori is based on a
project JPNP16007 commissioned by the New Energy and Industrial Technology
Development Organization (NEDO).
\end{acknowledgments}

\bibliography{quantum_computing}

\newpage
\onecolumngrid
\appendix
\section{Periodic boundary condition implementations}\label{app:pbc}
Recall that we work with charge neutral atoms and short-range (i.e.\ integrable) interatomic potentials with cutoffs.
Usually in such cases to calculate interaction terms with periodic boundary conditions, the
\textit{minimum image convention} is employed, in which the simulation cell is chosen
such that for any set of interacting atoms only one image of the
involved atoms should be within the cutoff distance of each other, so there is
a unique choice of which atoms interact~\cite{bulatov:2006aa}. This requires the unit cell to be
at least twice the size of the cutoff distance. As we cannot choose the cutoff distance
and the size of the required qubit number scales exponentially with the unit
cell size we cannot use the minimum image convention.

In this section we derive the direct sum formula for an $m$-body potential with
periodic boundary conditions and then show how to calculate the coefficients in
the HUBO in \cref{eqn:hamil}.

The energy of an infinite system due to an $m$-body potential $V_m$ with atoms
located on $x_1,x_2,\ldots \in \mathbb{R}^3$ is given as
\begin{align}
        \frac{1}{m!}\sum_{i_1\in\mathbb{N}}\sum_{\substack{i_2\in\mathbb{N}\\i_2\neq i_1}}
        \cdots \sum_{\substack{i_m\in \mathbb{N}\\i_m\neq i_1,\ldots, i_{m-1}}}
        V_m(x_{i_1}, \ldots, x_{i_m})\label{eqn:infinite_system}
\end{align}
Note that this includes the case where the atoms are of different species, for
which the actual parametrized form of $V_m$ would change depending on the
input and the case where we have periodic boundary conditions only on a subset
of basis vectors. We use the word atom on a location to mean an atom of a specific species
on a given location to simplify the notation from \cref{eqn:hamil}.

Assume now that the infinite system is generated by atoms on a unit cell on
locations $x_1, x_2,\ldots, x_N$ replicated following a set of lattice vectors
$\mathcal{L}$ so that \cref{eqn:infinite_system} becomes
\begin{align}\label{eqn:pbc_full_sum}
        \frac{1}{m!}
        \sum_{i\in {[N]}^m}
        \sideset{}{'}\sum_{\vec{n}_1,\ldots, \vec{n}_{m}\in \mathcal{L}}
        V_{m}(x_{i_1}+\vec{n}_1, \ldots,
        x_{i_m}+\vec{n}_m),
\end{align}
where we write $[N]:=\{1,\ldots, N\}$ and the prime on the sum indicates that
if $i=j$ then $\vec{n}_i \neq \vec{n}_j$, i.e.\ we exclude interactions with
two or more atoms on the same location. This sum can be interpreted as the
interaction terms of the unit cell given on $x_1+\vec{n}_1$ with the surrounding
super cell generated by the other lattice vectors. We thus define the energy of
a single unit cell by setting $\vec{n}_1=0$ as
\begin{align}\label{eqn:pbc_single_m}
        \frac{1}{m!}
        \sum_{i\in {[N]}^m}
        \sideset{}{'}\sum_{\vec{n}_2,\ldots, \vec{n}_{m}\in \mathcal{L}}
        V_{m}(x_{i_1}, x_{i_2}+\vec{n}_2,\ldots,
        x_{i_m}+\vec{n}_m),
\end{align}
where the prime condition on the sum is the same as before with $\vec{n}_1$ replaced by
$0$. For example for the two-body potential given by $q_iq_j/|r_i-r_j|$, where
$q_i$ and $q_j$ are the charges of the atoms on $x_i$ and $x_j$, we
recover the well-known formula~\cite{ewald:1921aa}
\begin{align}
        \frac{1}{2}
        \sum_{i\in {[N]}}\sum_{j\in [N]}
        \sideset{}{'}\sum_{\vec{n}\in \mathcal{L}}
        \frac{q_i q_j}{|x_i-x_j-\vec{n}|},
\end{align}
to calculate Coulomb interactions with periodic boundary conditions. For the
case with potentials of various order governing the system, e.g.
Stillinger-Weber with a two- and three-body part, we take the sum over $m$ to
obtain the total energy of a unit cell with periodic boundary conditions given
as
\begin{align}\label{eqn:pbc}
        E(\{x_1, x_2,\ldots, x_N\}):=\sum_{m\in [M]}
        \frac{1}{m!}
        \sum_{i\in {[N]}^m}
        \sideset{}{'}\sum_{\vec{n}_2,\ldots, \vec{n}_{m}\in \mathcal{L}}
        V_{m}(x_{i_1}, x_{i_2}+\vec{n}_2,\ldots,
        x_{i_m}+\vec{n}_m),
\end{align}
where $M$ is the highest order potential involved.

Let us now come to the calculation of the HUBO coefficients so that the sum over binary 
variables in \cref{eqn:hamil} reproduces \cref{eqn:pbc}. Consider a set of
lattice points $\{x_1,\ldots, x_m\}\subset X$ and associate to each point a
species so that we consider an atom of species $s_1$ on $x_1$ where $\{s_1,
s_2, \ldots, s_m\}$ is such that $s_i\in \mathcal{S}, i\in
[m]$. We define
\begin{align}\label{eqn:hubo_coeffs}
        H_{x_1, \ldots, x_m}^{s_1,\ldots, s_m}:=
        \sum_{\substack{\ell\in [M]\\\ell\ge m}}
        \frac{1}{\ell!}
        \sum_{\substack{i\in {[m]}^{\ell}\\ [m]\subset i}}
        \sideset{}{'}\sum_{\vec{n}_2,\ldots, \vec{n}_{\ell}\in \mathcal{L}}
        V_{\ell}(x_{i_1},
        x_{i_2}+\vec{n}_2,\ldots,
        x_{i_\ell}+\vec{n}_\ell)
\end{align}
where for simplicity we leave out the explicit writing of the species and the
condition $[m]\subset i$ on the second summation ensures that every index is
contained in $i$. This condition is needed to ensure that we only consider
potential contributions that require all the atoms and not only a subset which
would be part of a different HUBO coefficient.

To see that \cref{eqn:hubo_coeffs} is the correct way to define the HUBO
coefficients, we need to show that the sum in \cref{eqn:hamil} reproduces
\cref{eqn:pbc}. Let us consider a subset $\{y_1, \ldots, y_{N}\}=Y\subset X$
and a set $\{s_1, \ldots, s_{N}\}$ of species such that $b_{y_i}^{s_i}=1$
for $i\in [N]$ and $b_x^s=0$ else. The sum in \cref{eqn:hamil} then resolves to
\begin{align}
        \sum_{m\in [M]}\sum_{\vec{i}\in {[N]}^m}
        H_{x_{i_1}, \ldots, x_{i_m}}^{s_{i_1}, \ldots, s_{i_m}}
        &=\sum_{m\in [M]}\sum_{\vec{i}\in {[N]}^m}
        \sum_{\substack{\ell\in [M]\\\ell\ge m}}
        \frac{1}{\ell!}
        \sum_{\substack{j\in {\vec{i}}^{\ell}\\ \vec{i}\subset j}}
        \sideset{}{'}\sum_{\vec{n}_2,\ldots, \vec{n}_{\ell}\in \mathcal{L}}
        V_{\ell}(x_{j_1},
        x_{j_2}+\vec{n}_2,\ldots,
        x_{j_\ell}+\vec{n}_\ell)\\
        &=\sum_{\ell\in [M]}
        \frac{1}{\ell!}
        \sum_{\substack{m\in [M]\\m\le \ell}}\sum_{\vec{i}\in {[N]}^m}
        \sum_{\substack{j\in {\vec{i}}^{\ell}\\ \vec{i}\subset j}}
        \sideset{}{'}\sum_{\vec{n}_2,\ldots, \vec{n}_{\ell}\in \mathcal{L}}
        V_{\ell}(x_{j_1},
        x_{j_2}+\vec{n}_2,\ldots,
        x_{j_\ell}+\vec{n}_\ell),\label{eqn:check_pbc}
\end{align}
where the prime on the sum is in reference to the $j$ index, i.e.\ if
$j_{k}=j_{k'}$ then $\vec{n}_{k}\neq \vec{n}_{k'}$. Now use that the sums
$\sum_{\vec{i}\in {[N]}^m} \sum_{j\in {\vec{i}}^{\ell},
\vec{i}\subset j}$ can be written as the sum over all $\ell$-element multisets
with elements from $[N]$ that have exactly $m$ distinct elements, i.e.\ in an
abuse of notation we can write
\begin{align}
        \sum_{i\in {[N]}^m} \sum_{\substack{j\in {[i]}^{\ell}\\
        [i]\subset j}}=\sum_{j\in {[N]}^{\ell}} \mathbf{1}_{\text{$j$ has $m$ distinct
        elements}},
\end{align}
where $\mathbf{1}$ is the indicator function. Finally, since $j$ has $\ell$
elements we have
\begin{align}
        \sum_{\substack{m\in [M]\\m\le \ell}}
        \mathbf{1}_{\text{$j$ has $m$ distinct elements}}
        = 1
\end{align}
and thus \cref{eqn:check_pbc} can be written as
\begin{align}
        \sum_{\ell\in [M]}
        \frac{1}{\ell!}
        \sum_{j\in {[N]}^{\ell}}
        \sideset{}{'}\sum_{\vec{n}_2,\ldots, \vec{n}_{\ell}\in \mathcal{L}}
        V_{\ell}(x_{j_1},
        x_{j_2}+\vec{n}_2,\ldots,
        x_{j_\ell}+\vec{n}_\ell),
\end{align}
and we recovered \cref{eqn:pbc}.

There is an efficient way to calculate
\cref{eqn:hubo_coeffs} when you have access to an oracle that calculates the
total energy as is for example the case in the OpenKIM API. This oracle
for atoms on some locations $Y=y_1, \ldots, y_N\in \mathbb{R}$ returns
\begin{align}
        F_{\ell}(Y) :=
        \sum_{\substack{i\in {[N]}^\ell\\ i_1<
        i_2<\ldots< i_\ell}}
        V_\ell(y_{i_1}, y_{i_2}, \ldots, y_{i_\ell})=
        \frac{1}{\ell!}
        \sideset{}{''}\sum_{i\in {[N]}^\ell}
        V_\ell(y_{i_1}, y_{i_2}, \ldots, y_{i_\ell}),
\end{align}
where again we leave out the explicit mention of the species on the potential,
use that the potential is constant under permutation of arguments and the
double prime indicates that no two indices $i_k$, $i_{k'}$ should be the same
in the summation
(this is to simplify the notation from \cref{eqn:infinite_system}). Recall that
the potentials that we use have a hard cutoff. To calculate $H_{x_1, \ldots,
x_m}^{s_1, \ldots, s_m}$ construct a super cell by adding copies of the
configuration in the unit cell around the unit cell in the directions in which
we have periodic boundary conditions up until the atoms in the unit cell have
no non-zero interaction with the newly copied unit cells. As an example, for
the MoS$_2$ system this means that we create a $5\times 5$ cell of unit cells
with the copied configurations. Call this set $SC$ and their elements $y_1,y_2,
\ldots, y_{|SC|}$ and note that the set $\mathcal{L}$ of lattice vectors is
given by the basis vectors of the unit cell. Now,
\begin{align}
        F_{\ell}(SC) - F_{\ell}(SC\setminus \{x_j\})
        &=
        \frac{1}{\ell!}
        \sideset{}{''}\sum_{\substack{z_1,\ldots, z_\ell\in SC\\ \exists
        k\in[\ell]: z_k = x_j}}
        V_\ell(z_1, z_2, \ldots, z_\ell)\\
        &=
        \frac{1}{(\ell-1)!}
        \sideset{}{''}\sum_{z_2,\ldots, z_\ell\in SC}
        V_\ell(x_j, z_2, \ldots, z_\ell)
\end{align}
so that
\begin{align}
        \sum_{j\in [m]} F_{\ell}(SC) - F_{\ell}(SC\setminus \{x_j\})
        &=
        \frac{1}{(\ell-1)!}
        \sum_{i\in {[m]}^{\ell}}
        \sideset{}{'}\sum_{\vec{n}_2,\ldots, \vec{n}_{\ell}\in \mathcal{L}}
        V_\ell(x_{i_1}, x_{i_2}+\vec{n}_2, \ldots, x_{i_\ell}+\vec{n}_\ell),
\end{align}
where we used again that the potential is constant under permutation of
arguments. The configuration energy with periodic boundary conditions
\cref{eqn:pbc} is thus obtained by
\begin{align}
        E(UC)=\sum_{\ell\in [M]}
        \frac{1}{\ell}\sum_{j\in [m]}\left[ F_{\ell}(SC) - F_{\ell}(SC\setminus
        \{x_j\})\right].
\end{align}
We can now calculate the linear HUBO coefficients in \cref{eqn:hubo_coeffs} as
\begin{align}
        H_{x}^t = \sum_{\ell\in [M]} \frac{1}{\ell!}
        \sideset{}{'}\sum_{\vec{n}_2,\ldots, \vec{n}_{\ell}\in \mathcal{L}}
        V_{\ell}(x, x+\vec{n}_2, \ldots, x+\vec{n}_{\ell})= E(\{x\}).
\end{align}
Now, for quadratic terms we find
\begin{align}
        H_{x_1,x_2}^{s_1, s_2} =  E(\{x_1, x_2\}) - E(\{x_1\}) - E(\{x_2\})
\end{align}
which is easily seen by looking at the second sum in \cref{eqn:hubo_coeffs}
which considers any multiset of indices that contains the entirety of the
original set, i.e.\ here $\{1, 2\}$ and by subtracting the single atom energies
on the right-hand side, we subtract those contributions that arise from the
summands in which only a single index, either $1$ or $2$ is present. It is now
clear how to generalise this
\begin{align}
        H_{x_1,\ldots, x_m}^{s_1,\ldots, s_m} =  E(\{x_1, \ldots, x_m\}) -
        \sum_{Y\subsetneq \{x_1, \ldots, x_m\}} H_{Y}^{s_Y}
\end{align}
where on the right hand side we write $H_Y^{s_Y}$ for the coefficient with
atoms on positions given by $Y$ and the appropriate species set $s_Y$.

We close this appendix with a remark on non-parametrized potentials in which
you do not have access to the $n$-body potential part separately so that
the oracle \cref{eqn:oracle} looks like
\begin{align}\label{eqn:oracle}
        F(Y) :=
        \sum_{\ell\in [M]}\frac{1}{\ell!}
        \sideset{}{''}\sum_{i\in {[N]}^\ell}
        V_\ell(y_{i_1}, y_{i_2}, \ldots, y_{i_\ell}).
\end{align}
In this case we have
\begin{align}
        F(SC) - F(SC\setminus \{x_j\})
        &=
        \sum_{\ell\in [M]}
        \frac{1}{(\ell-1)!}
        \sum_{i\in {[m]}^{\ell-1}}
        \sideset{}{'}\sum_{\vec{n}_2,\ldots, \vec{n}_{\ell}\in \mathcal{L}}
        V_\ell(x_j, x_{i_2}+\vec{n}_2, \ldots, x_{i_\ell}+\vec{n}_\ell),
\end{align}
and thus it is not clear whether there exists an efficient algorithm to calculate
$E(UC)$ with such an oracle.

\section{Cohesive energy}\label{app:cohesive}
When doing grand canonical calculations we need to ensure that the energies
with different numbers of atoms are comparable. We use the notion of cohesive
energy for this, which is usually defined as the difference in energy between
the atoms in a specific configuration and the energy of all the involved atoms
at an infinite pairwise distance. In our case this means that we compare the
energy of a configuration on the lattice with the regular lattice constant $a$
and the energy with $a\rightarrow \infty$. For these energy calculations we use
interatomic potentials with a hard cutoff and
thus the energy of the atoms with an infinite pairwise distance is $0$ while it
is non-zero for the regular lattice constant. Thus the cohesive energy in our
case is calculated by \cref{eqn:hamil} as claimed in the main text.
\section{Local minima}\label{app:loc_mins}
\begin{figure}
\centering
\includegraphics[width=0.5\textwidth]{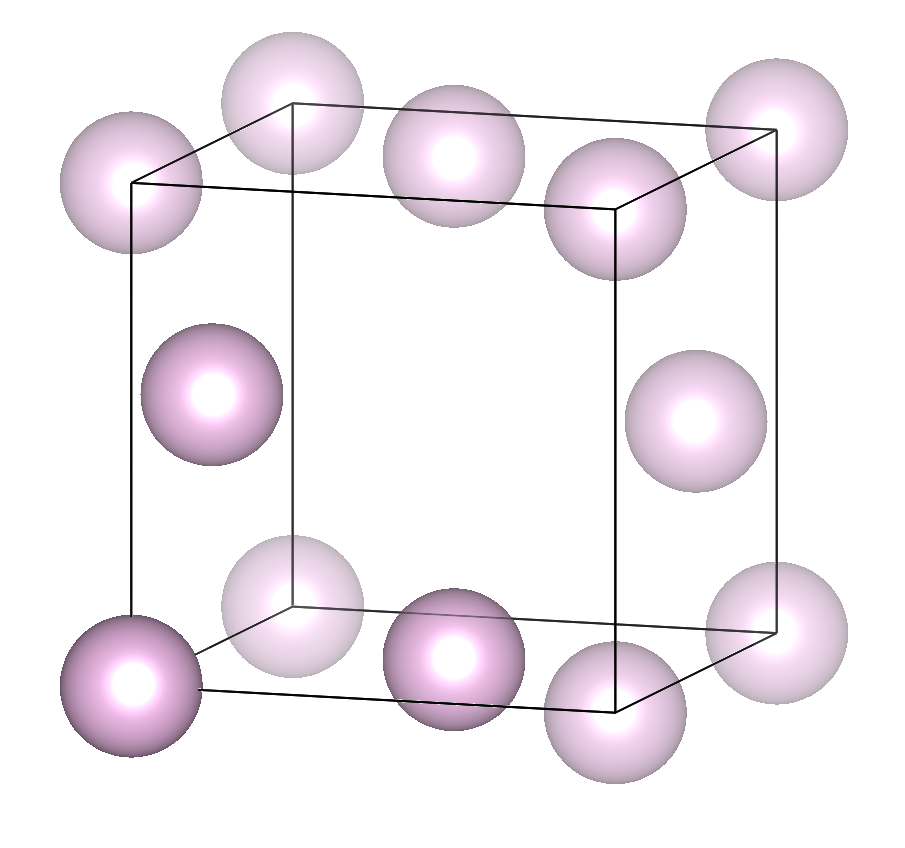}
\caption{Kr$_3$ configuration that corresponds to an FCC configuration with a
single atom taken out and which has a residual energy of
$0.2029$eV}\label{fig:lj_loc_mins}
\end{figure}

\begin{figure}
\centering
\subcaptionbox{}{\includegraphics[width=0.3\textwidth]{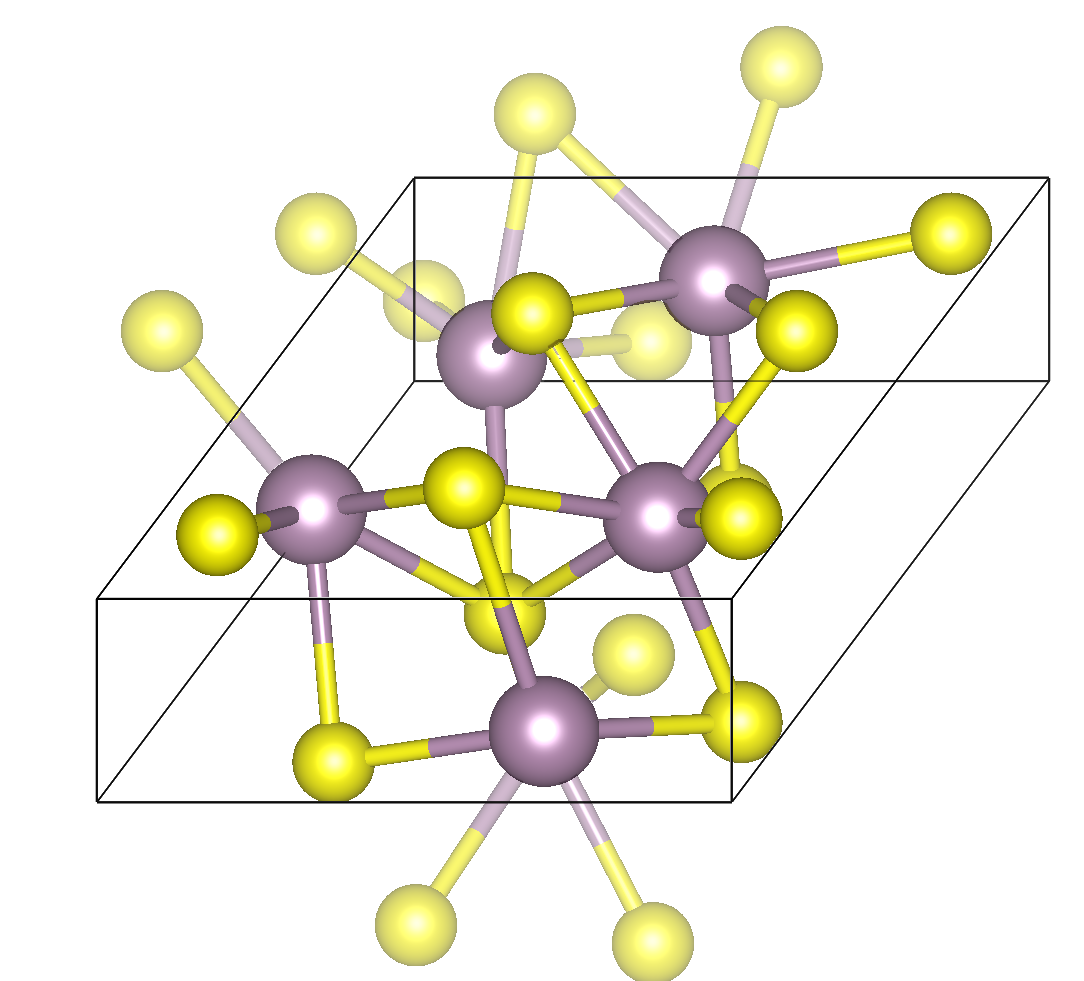}}%
\hfill
\subcaptionbox{}{\includegraphics[width=0.3\textwidth]{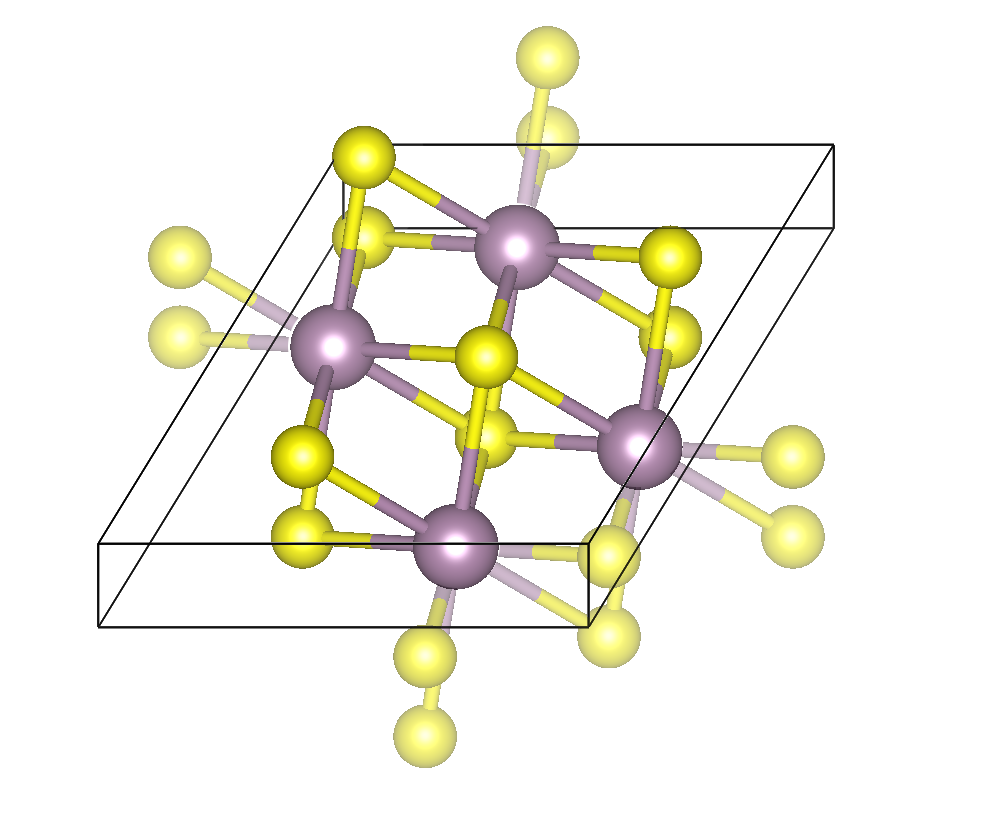}}%
\hfill
\subcaptionbox{}{\includegraphics[width=0.3\textwidth]{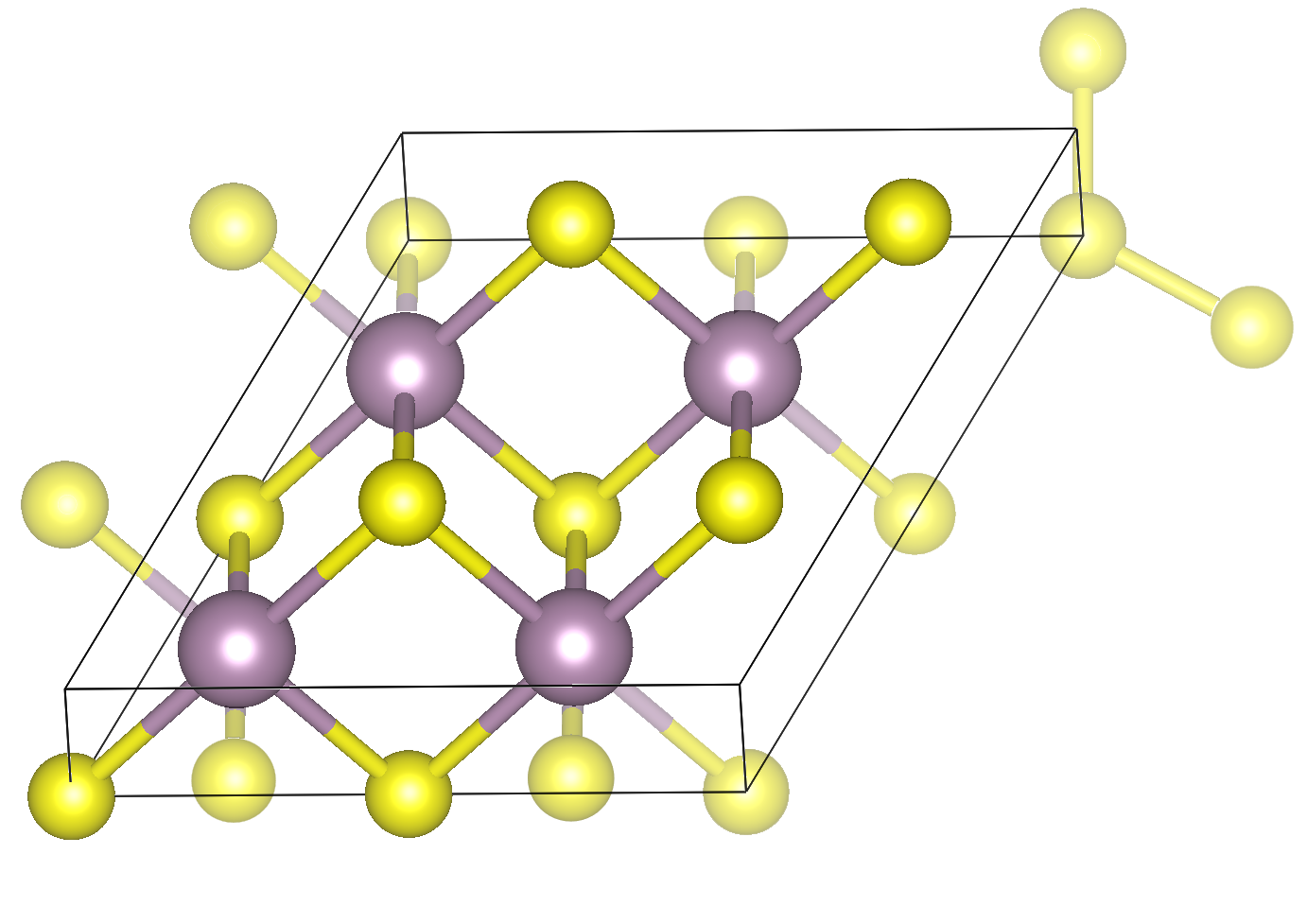}}%
\caption{Local minima of the MoS$_2$ system marked with a dotted line in
        \cref{fig:mos2_histogram}. From left to right, (a) an example Mo$_5$S$_{10}$
        configuration with a residual energy of $-6.2161$eV, (b) the
        orthorombic state with a residual energy of $-0.9313$eV and (c) the
$1\mathrm{T}$ configuration of MoS$_2$ $1.4755$eV}\label{fig:mos2_loc_mins}
\end{figure}

We give an overview of the local minima indicated by dotted lines in the
histograms \cref{fig:lj_histogram,fig:mos2_histogram}. The local minima for
the Krypton system are given in \cref{fig:lj_loc_mins} and for the MoS$_2$ system in
\cref{fig:mos2_loc_mins}.
\end{document}